\documentclass[conference]{IEEEtran}
\IEEEoverridecommandlockouts
\usepackage{cite}
\usepackage{float}
\usepackage{amsmath,amssymb,amsfonts}
\usepackage{graphicx, enumitem}
\usepackage{textcomp}
\usepackage[ruled,vlined,linesnumbered]{algorithm2e}
\usepackage[table,dvipsnames]{xcolor}
\usepackage{tikz}
\usepackage{xcolor}
\usepackage{soul}
\usepackage{tabularx}
\usepackage{bm}
\usepackage{makecell}
\usepackage{booktabs}
\usepackage{multirow}
\usepackage{url}
\usepackage{subfigure}
\usepackage{tikz}
\usepackage[none]{hyphenat}
\SetKwIF{If}{ElseIf}{Else}{if}{}{else if}{else}{end if}%
 
\setlist[enumerate]{leftmargin=15mm}
\newcommand\crule[1][black]{\textcolor{#1}{\rule{0.2cm}{0.2cm}}}

\definecolor{LightBlue}{RGB}{0,110,255}
\definecolor{Yellow}{RGB}{247,200,47}
\definecolor{Green}{RGB}{179,200,23}
\definecolor{Green2}{RGB}{0,204,0}
\definecolor{Purple}{RGB}{126,26,200}
\definecolor{Red}{RGB}{255,0,0}
\definecolor{GreenLight}{RGB}{50,205,50}
\definecolor{Black}{RGB}{0,0,0}
\definecolor{Brown}{RGB}{204,102,0}

\makeatletter
\newcommand*\titleheader[1]{\gdef\@titleheader{#1}}
\AtBeginDocument{%
  \let\st@red@title\@title
  \def\@title{%
    \bgroup\normalfont\large\centering\@titleheader\par\egroup
    \vskip0.5em\st@red@title}
}
\makeatother

\definecolor{GREEN}{RGB}{0, 140, 80}

\renewcommand{\theenumi}{\Alph{enumi}}

\newcolumntype{L}[1]{>{\raggedright\let\newline\\\arraybackslash\hspace{0pt}}m{#1}}
\newcolumntype{C}[1]{>{\centering\let\newline\\\arraybackslash\hspace{0pt}}m{#1}}
\newcolumntype{R}[1]{>{\raggedleft\let\newline\\\arraybackslash\hspace{0pt}}m{#1}} 

\def\BibTeX{{\rm B\kern-.05em{\sc i\kern-.025em b}\kern-.08em
    T\kern-.1667em\lower.7ex\hbox{E}\kern-.125emX}}

\makeatother

\begin{document}

\title{Optimizing Package Delivery with Quantum Annealers: Addressing Time-Windows and Simultaneous Pickup and Delivery\\
	{}
	\thanks{This work was supported by the Basque Government through Plan complementario comunicación cuántica (EXP. 2022/01341) (A/20220551) and HAZITEK program (QUANTHIB project, ZL-2024/00334). During the preparation of this work the authors used Microsoft Copilot in order to improve the language and readability of the manuscript. After using this tool/service, the authors reviewed and edited the content as needed and takes full responsibility for the content of the publication.}
}
\author{
	\IEEEauthorblockN{Eneko Osaba\IEEEauthorrefmark{2}\IEEEauthorrefmark{1}, 
        Esther Villar-Rodriguez\IEEEauthorrefmark{2},
        Pablo Miranda-Rodriguez\IEEEauthorrefmark{2} and
        Antón Asla\IEEEauthorrefmark{3}}
	\IEEEauthorblockA{\IEEEauthorrefmark{2}TECNALIA, Basque Research and Technology Alliance (BRTA), 48160 Derio, Spain}
    \IEEEauthorblockA{\IEEEauthorrefmark{3}Serikat - Consultoría y Servicios Tecnológicos, 48009 Bilbao, Spain}
    \IEEEauthorblockA{\IEEEauthorrefmark{1}Corresponding author. Email: eneko.osaba@tecnalia.com}}

\maketitle

\IEEEpubidadjcol	

\begin{abstract}
Recent research at the intersection of quantum computing and routing problems has been highly prolific. Much of this work focuses on classical problems such as the Traveling Salesman Problem and the Vehicle Routing Problem. The practical applicability of these problems depends on the specific objectives and constraints considered. However, it is undeniable that translating complex real-world requirements into these classical formulations often proves challenging. In this paper, we resort to our previously published quantum-classical technique for addressing real-world-oriented routing problems, known as \textit{Quantum for Real Package Delivery} (\texttt{Q4RPD}), and elaborate on solving additional realistic problem instances. Accordingly, this paper emphasizes the following characteristics: \textit{i)} simultaneous pickup and deliveries, \textit{ii)} time-windows, and \textit{iii)} mobility restrictions by vehicle type. To illustrate the application of \texttt{Q4RPD}, we have conducted an experimentation comprising seven instances, serving as a demonstration of the newly developed features.
\end{abstract}


\section{Introduction}\label{sec:intro}

The Package Delivery Problem (PDP) is a combinatorial optimization problem that involves efficiently routing and scheduling the pickup and delivery of a set of packages to various destinations while minimizing cost and travel time. The main objectives of the PDP are \textit{i)} to pickup and deliver all packages and \textit{ii)} to distribute deliveries among a fleet of vehicles, while doing so using the most cost-effective routes.

The PDP is an interesting problem for the operations research community, since it has a wide range of applicability, especially in industry \cite{hanif2015applying,benarbia2021literature}. It is also a NP-hard problem, so even for relatively small problems, current classical computers find brute-force methods challenging. Thus, because of their significant scientific and business interest, PDP variants have been extensively studied in the literature using various solving approaches. In this paper, we examine the PDP from a perspective that has been relatively underexplored: Quantum Computing (QC). Currently, QC is attracting considerable attention from researchers and practitioners, with the community striving to identify the contributions that this groundbreaking technology can offer to numerous application sectors. Transportation and logistics are areas where QC has already shown promising results \cite{osaba2022systematic}. Despite this growing scientific interest, the PDP variants have been relatively underexamined from the QC perspective, where most of the routing research has been carried out around the canonical formulation of the Traveling Salesman Problem (TSP, \cite{gutin2006traveling}) and the Vehicle Routing Problem (VRP, \cite{toth2002vehicle}).

In this context, our primary goal is to advance beyond our previous study presented in \cite{osaba2024solving}. In that study, we tackled a problem of goods distribution with a heterogeneous fleet using a specially developed QC-based method. This hybrid approach allows for the classification of certain customers as priority, requiring service before a specified time threshold. In addition, the problem accommodates customers with multiple orders, which can be fulfilled by more than one vehicle. We named this solver \textit{Quantum for Real Package Delivery} (\texttt{Q4RPD}). In this paper, we expand upon the formulation and findings published in \cite{osaba2024solving} by addressing the following aspects:
\begin{itemize}
    \item To broaden the scope of customer demands, \textit{we now enable customers to request the collection of certain materials in addition to deliveries}. This functionality is particularly significant in contexts such as pharmacological waste collection and recycling of used glass containers. It is important to highlight that our system accommodates customers who request only pickup, only delivery, or simultaneous pickup and delivery.
    
    \item To enhance the practicality of \texttt{Q4RPD}, we now incorporate one of the most commonly used features in real-world logistics: \textit{time-windows} \cite{golden2008vehicle}. This constraint specifies the time intervals within which certain tasks must be completed.
    
    \item To further enhance the characteristic of the heterogeneous vehicle fleet, \textit{we now enable the imposition of specific mobility restrictions by vehicle type}. In real-world scenarios where deliveries span various urban areas, not all vehicles in a fleet may be able to access the locations of the customers. For instance, it is common to encounter city zones where certain trucks are restricted from entering. To address this issue, we enable the characterization of each order to determine which vehicles in the fleet can visit them.

\end{itemize}

All these new features, integrated into \texttt{Q4RPD}, facilitate the delineation of more realistic problem instances, thereby enhancing their applicability to the field of transportation and logistics. Additionally, we perform experiments to deal with real-oriented instances in new scenarios that were not considered in the original formulation, demonstrating the flexibility of our method to handle a broader spectrum of use cases.

It should be noted that all the advances described in this research were developed in collaboration with Ertransit\footnote{https://www.ertransit.com.cn/ertransit-espana/}, a Spanish company that specializes in transport and logistics. A series of meetings between Ertransit and the research team led to the identification of real-world needs, which subsequently guided the developments detailed in this study.

The remainder of this paper is organized as follows: Section \ref{sec:back} provides the necessary background to contextualize our work. Section \ref{sec:Q4RPD} describes \texttt{Q4RPD} and all the newly implemented functionalities. In Section \ref{sec:demo}, these new functionalities are demonstrated through an experimentation featuring seven instances. Finally, Section \ref{sec:conc} is dedicated to conclusions and future work.

\section{Background}\label{sec:back}
In general, the Vehicle Routing Problem (VRP) is an umbrella that covers a myriad of different problems of industrial and logistical interest, from which the Capacitated Vehicle Routing Problem (CVRP) is typically considered the most basic one \cite{dantzig1959truck}. In VRPs, the key decisions revolve around allocating customers to specific vehicles and planning the order in which each vehicle will visit its assigned customers. The problem is analogous to several instances of the Traveling Salesman Problem (TSP) if this allocation is done beforehand \cite{mor2022vehicle}. Since there are several taxonomies of the VRPs \cite{EKSIOGLU20091472, braekers2016vehicle}, in this work we call Package Delivery Problem to the family of the sometimes called VRP with Pickup Delivery (VRPPD) and its extensions, such as VRP with backhauls (VRPB) \cite{braekers2016vehicle}. In our definition, the core part of the PDP is the fact that goods (which have weight and volume) need to be picked and delivered at several locations. 

Since this is a recurrent industry problem, various time-efficient solving techniques have been developed in classical computing over the past few decades. These include exact methods \cite{chinan1}, heuristics \cite{heuristic2}, and metaheuristics \cite{parcel}. Other advanced approaches such as Reinforcement Learning \cite{zhang2021solving} and Deep Learning \cite{xin2020step} have also been explored to address challenging routing problems.

QC is a paradigm that leverages the properties of quantum mechanics to perform computations using quantum bits or qubits. This novel premise has been explored in many different areas, ranging from fundamental physics, simulations, cryptography, machine learning, to combinatorial optimization. Considering this last field, while there is little hope about QC algorithms solving NP-complete problems efficiently \cite{altshuler2010anderson}, there is evidence that quantum algorithms could outperform classical ones for some specific problems \cite{somma2012quantum, hastings2021power, tasseff2024emerging}.

Regarding the interplay between PDP and QC, as stated before, most of the work has been developed around the more general problem VRP. Indeed, there are many studies on solving the VRP with different QC algorithms, such as QAOA \cite{azad2022solving} and QSVM \cite{mohanty2024solving} (gate-based QC), or Quantum Annealing \cite{tambunan2023quantum} (analog QC). Nevertheless, there is also state of the art research that focuses on solving complex and realistic routing problems, such as \cite{weinberg2023supply}, where multi-truck vehicle routing is solved by iteratively assigning each truck to a route, or \cite{cattelan2024modeling}, where multiple customers can request on‑demand pickups and drop‑offs from shared vehicles within a fleet. 

The framework \texttt{Q4RPD}, introduced in our previous work \cite{osaba2024solving}, could also be included in this category since it addressed delivery priorities, heterogeneous fleets, and two-dimensional descriptions of items. The present work is an extension of the former framework that goes beyond the original constraints and aims to consider further real-world restrictions to model what we call PDP.

\section{Quantum Framework for Real-World Package Delivery - \texttt{Q4RPD}}\label{sec:Q4RPD}

In a nutshell, \texttt{Q4RPD} is a quantum-based method that can deliver highly effective solutions to real-world logistic problems by leveraging the advanced capabilities of current quantum devices. Throughout the execution of the solver, (sub-)routes are iteratively assigned to the available vehicles while adhering to all existing restrictions. \texttt{Q4RPD} amalgamates both quantum and classical computing as follows:

\begin{itemize}
    \item \textit{Classical computing} is employed to oversee the general workflow of the algorithm and to execute processes such as the intelligent allocation of resources and the integration of subsolutions provided by the quantum annealer.
        
    \item \textit{Quantum computing} is utilized for building each (sub-)route to be added into the complete solution. To construct each route, \texttt{Q4RPD} employs D-Wave's Leap Constrained Quadratic Model (CQM) Hybrid Solver (\texttt{LeapCQMHybrid}\cite{leapCQM}).    
\end{itemize}

\noindent\begin{table*}[t]
\centering
\caption{The complete set of parameters and variables used in our formulation.}\label{tab:params_vars}
\resizebox{1.0\textwidth}{!}{
    \begin{tabular}{lp{13.0cm}}
    \hline
    \makecell[l]{\cellcolor{gray!10}\textbf{Upper bound} \\\cellcolor{gray!10}\textbf{parameters}} & \cellcolor{gray!10} \\ 
    \cellcolor{gray!10}$rt$ & \cellcolor{gray!10}the maximum possible duration of a (sub-)route. \\
    \cellcolor{gray!10}$W,D$ & \cellcolor{gray!10}the permissible maximum weight and dimension in the truck assigned to the (sub-)route. \\
    \cellcolor{gray!10}$vt$ & \cellcolor{gray!10}the type of truck assigned to the (sub-)route. \\[2mm]
    \makecell[l]{\textbf{Static} \\ \textbf{parameters}}\\
    P & set of orders still to be met.\\
    $M$ & number of orders (cardinality of P). \\
    $wd_i, dd_i$ & weight and dimension of material delivered in order $i\in$ P. \\
    $wp_i, dp_i$ & weight and dimension of material picked up in order $i\in$ P. \\
    $LT, UT$ & sets of lower and upper time limits of the orders, respectively.\\
    $lt_i, ut_i$ & the lower and upper time limits for order $i$. \\
    $ot_i$ & a numerical value to describe the geographical area of the client assisgned to order $i$. \\
    $c_{i,j}$, $d_{i,j}$, & travel time and distance associated with going from the location of order $i$ to the location of order $j$. \\[2mm]
    \multicolumn{1}{l}{\bf Variables} \\
    $x_{i,p}$ & binary variable that represents if the location of order $i$ is visited at position $p$ of the (sub-)route, with $p \in [0,M]$. \\ \hline
    \end{tabular}
}
\end{table*}%

\begin{figure}[h]
    \centering
    \includegraphics[width=1.0\linewidth]{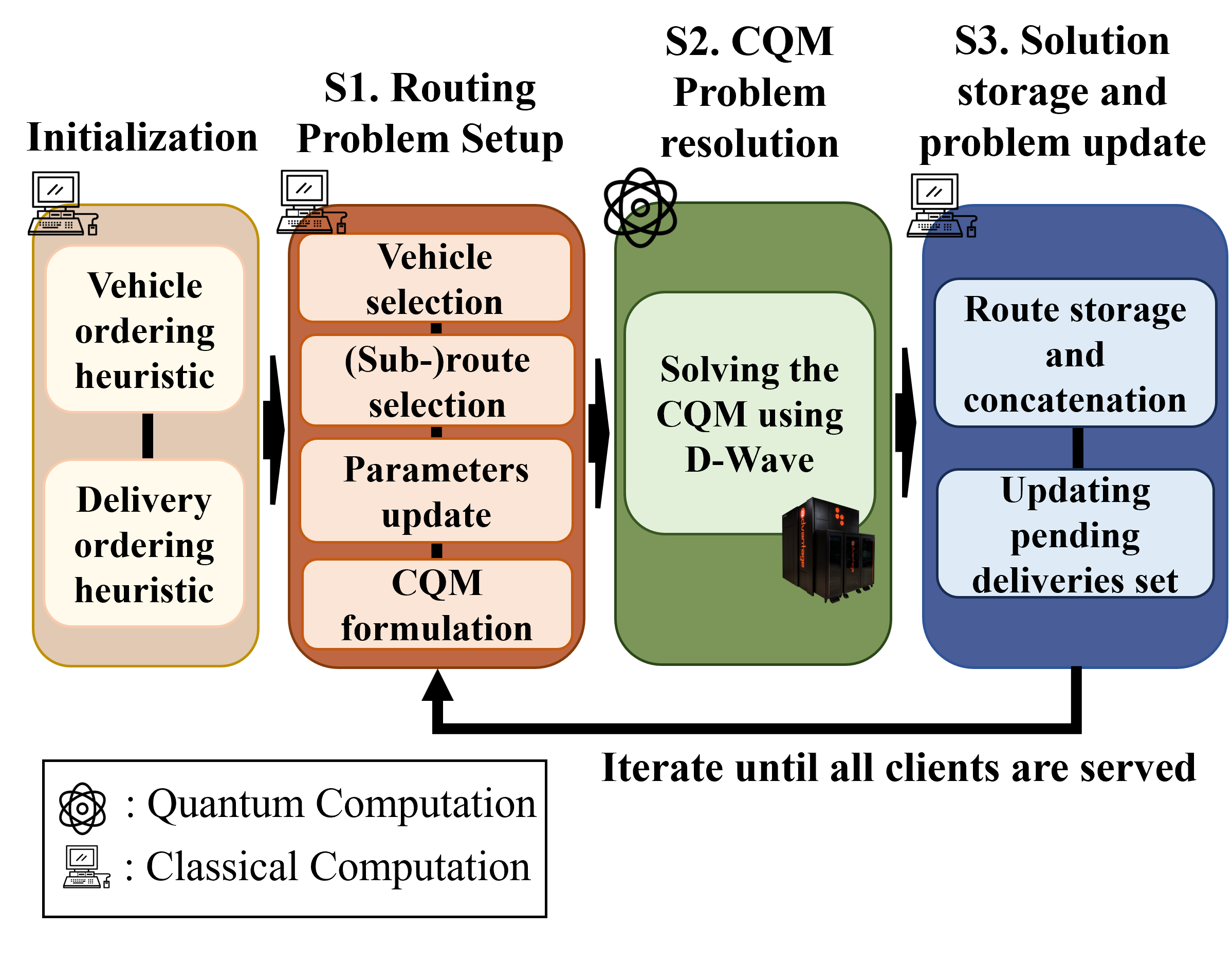}
    \caption{General workflow of \texttt{Q4RPD}.}
    \label{fig:Q4RPD}
\end{figure}

\texttt{Q4RPD} is able to calculate four types of (sub-)routes, categorized by their origin and destination. These categories arise from the stringent restrictions imposed by the Top-Priority (TP) order\footnote{Previously known as Top-Priority \textit{delivery}, but redefined as \textit{order} to encompass both delivery and pickup}, which must be satisfied before a specified deadline. The four types of routes that \texttt{Q4RPD} can calculate are \textit{Depot-\texttt{TP} sub-route}, \textit{\texttt{TP}-\texttt{TP} sub-route}, \textit{\texttt{TP}-Depot sub-route}, and \textit{Regular route}. Nonetheles, as this paper aims to showcase the newly implemented features, we will exclude TP orders from all experimental instances. Therefore, all calculated paths are classified as \textit{regular}, which are complete paths that begin and finish at the depot.



For the sake of thoroughness, we depict the main workflow of \texttt{Q4RPD} in Fig. \ref{fig:Q4RPD}. For further details, we refer interested readers to \cite{osaba2024solving}. Thus, the remainder of this section outlines the newly implemented functionalities. The parameters and variables used are detailed in Table \ref{tab:params_vars}.


\textbf{Simultaneous Pickup and Delivery:} in a logistics and industrial environment where recycling is gaining more and more importance, it is crucial that route planning systems not only account for the delivery of materials, but also be equipped for the collection of items from customers. In our previous system, by using parameters ($wd_i,dd_i$) (detailed in Table \ref{tab:params_vars}), we enabled customers to request the delivery of specific materials. In this new version of \texttt{Q4RDP}, we enhance this functionality by introducing parameters ($wp_i,dp_i$), which denote the weight and dimensions of the load that the customer wishes to be picked up. It is important to note that our system allows customers to request either the delivery or collection of materials, or both simultaneously. 

It is also worth mentioning that, as can be seen in the recent survey published in \cite{phillipson2024quantum}, simultaneous pickup and delivery has never been tackled by any quantum-based solver before. This contribution, therefore, adds significant originality to our research. Consequently, to properly integrate this new functionality, a new set of restrictions must be established to ensure that at each position $p$ of the route, the sum of the material collected up to that point and the items yet to be delivered does not exceed the vehicle's maximum capacity. Thus, for each position $p$, a pair of constraints is generated such that:




\begin{itemize}
    \item \textbf{Weight-pickup-restriction-at-$p$}: the total of all pickups made and deliveries yet to be completed must not exceed the vehicle's maximum allowed weight $W$. 
    \begin{equation}\label{eq:weight_p}
    \sum_{p'=0}^{p}x_{i,p'}wp_i + \sum_{p'=p+1}^{M}x_{i,p'}wd_i \leq W \quad\forall i\in \{0,\dots,M\}.
    \end{equation}
    \item \textbf{Dimension-pickup-restriction-at-$p$}: the total of all pickups made and deliveries yet to be completed must not exceed the vehicle's maximum allowed dimension $D$. 
    \begin{equation}\label{eq:dimension_p}
    \sum_{p'=0}^{p}x_{i,p'}dp_i + \sum_{p'=p+1}^{M}x_{i,p'}dd_i \leq D \quad\forall i\in \{0,\dots,M\}.
    \end{equation}
\end{itemize}

\textbf{Time Windows:} This type of constraint is undoubtedly the most extensively studied from a classical perspective. The significance of time windows in a logistics environment is unquestionable, given that orders must be fulfilled for various businesses or individuals, each with their specific schedules and availabilities. Whereas time windows have been theoretically or preliminarily addressed from a quantum perspective in studies such as \cite{irie2019quantum} and \cite{harwood2021formulating}, their application to medium or large-sized problems has remained largely unexplored \cite{weinberg2023supply}.

In the previous version of \texttt{Q4RPD}, the upper time limit for a TP order was established employing the $ut_i$ parameter (formerly known as $t_i$). In this new version of \texttt{Q4RDP}, we extend our formulation by introducing a lower time limit parameter $lt_i$, hence creating a time window ($lt_i, ut_i$). This apparently minor change significantly increases the complexity of the problem and its formulation, as the new version forces all customers to be visited within their designated time windows. This requires the introduction of a set of constraints that ensure, at each position on the route, the strict adherence to the lower and upper time limits of the visited customer. As a consequence, for each position $p$, a pair of restrictions is generated such that:

\begin{itemize}
    \item \textbf{Lower-time-limit-at-$p$}: the route's duration at $p$ must be longer than the lower time window of the visited location. 
    \begin{equation}\label{eq:lower_time}
        \begin{aligned}
            \sum_{i=0}^{M}\sum_{j=0}^{M}c_{i,j}x_{i,p'}x_{j,p'+1} \geq  \sum_{k=0}^{M}LT_k*x_{k,p} \\ \quad\forall p'\in \{0,\dots,p-1\}, i\neq j.
        \end{aligned}
    \end{equation}
    \item \textbf{Upper-time-limit-at-$p$}: the route's duration at $p$ must not be longer than the upper time window of the visited location.
    \begin{equation}\label{eq:upper_time}
        \begin{aligned}
            \sum_{i=0}^{M}\sum_{j=0}^{M}c_{i,j}x_{i,p'}x_{j,p'+1} \leq \sum_{k=0}^{M}UT_k*x_{k,p} \\ \quad\forall p'\in \{0,\dots,p-1\}, i\neq j.
        \end{aligned}
    \end{equation}
\end{itemize}

Being $LT$ and $UT$, as described in Table \ref{tab:params_vars}, two ordered sets containing all the lower and upper limits of the orders, respectively. Finally, it is worth noting that \texttt{Q4RPD} allows for the existence of customers with partial or non-existent time windows, through $lt_i = 0$ and/or $ut_i = \infty$.

\textbf{Mobility restrictions by vehicle type:} Logistics companies are required to serve customers in different geographical areas on a daily basis. Accordingly, fleet vehicles must travel to customers in rural areas, city outskirts, and densely populated urban centers. This can lead to certain limitations, as some vehicles might be restricted from entering specific urban areas due to factors like size or emissions.

\begin{figure}[t]
    \centering
    \includegraphics[width=0.45\linewidth]{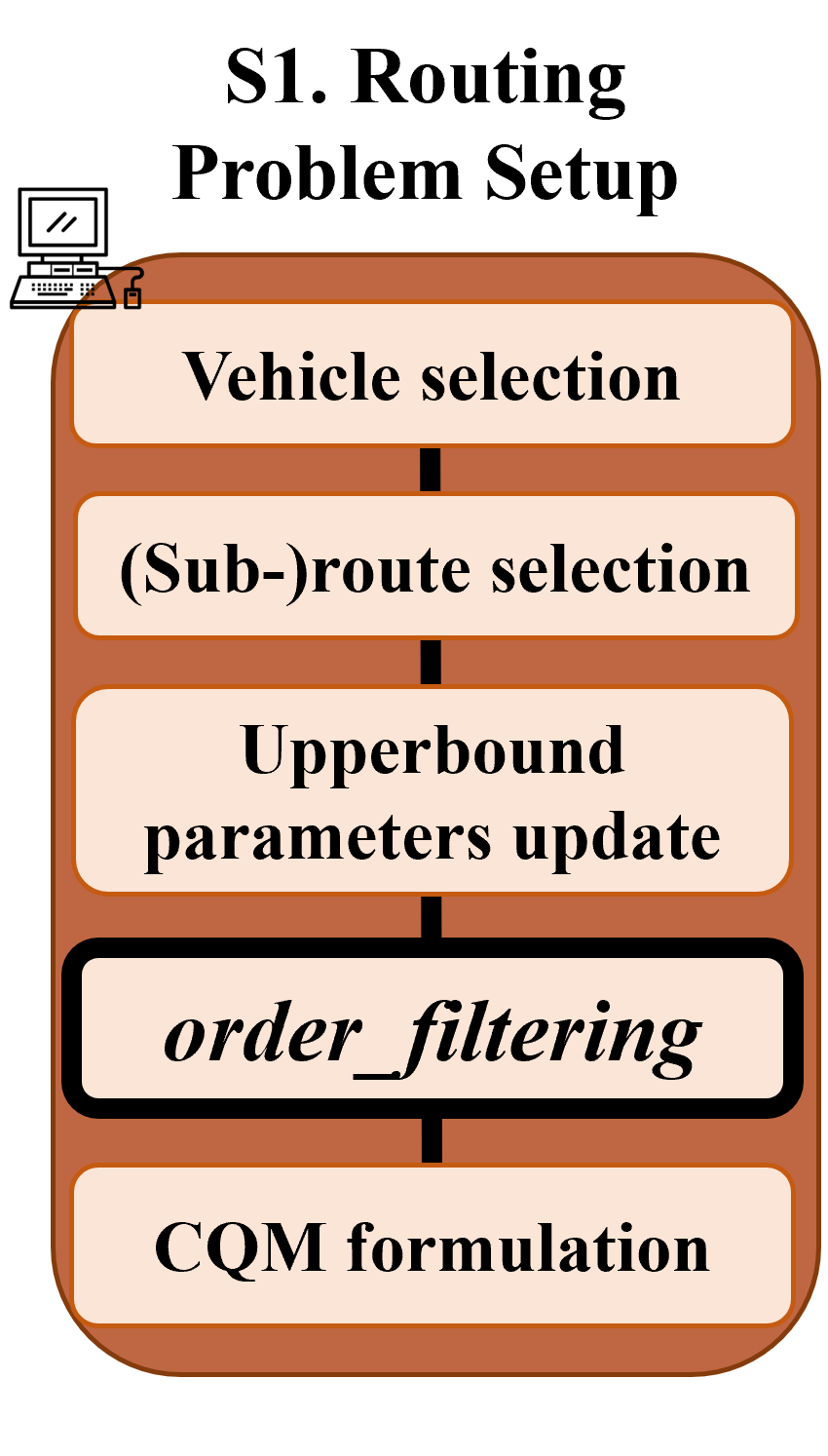}
    \caption{New workflow of S1. Routing Problem Setup.}
    \label{fig:newS1}
\end{figure}

In this study, we have expanded the application of \texttt{Q4RPD} to consider this real-world scenario. To do this, we have introduced a parameter $vt$ to describe the category of the trucks, and $ot_i$ to specify the type of area where the customer associated to order $i$ is located. Both parameters are assigned an integer numerical value, ensuring that a vehicle $i$ can only fulfill an order if $vt \leq ot_i$. For instance, a vehicle with $vt$=3 can only visit customers with $ot_i\geq3$, whereas a vehicle with $vt$=1 can serve all orders.

This real-world situation can be managed in various ways by \texttt{Q4RPD}. One possible method involves extending the mathematical formulation of the CQM, ensuring that an order $i$ is included in the route only if its type is compatible with the assigned vehicle. That is:

\begin{equation}\label{eq:MR1}
    \begin{aligned}
        \sum_{p=0}^{M} x_{i,p} \leq \begin{cases} 
        1 & \text{if } vt \leq ot_i \\
        0 & \text{otherwise}
        \end{cases} \quad \forall i \in \{0, \dots, M\}
    \end{aligned}
\end{equation}

Although this approach is valid, it significantly increases the problem's complexity by adding new constraints. To mitigate this issue, a classical filtering mechanism called \texttt{order\_filtering} has been implemented. This procedure examines the type of the assigned vehicle and creates a subset $P'\subset P$, including only the orders that the vehicle can handle based on its type. An additional benefit of this approach is that it may reduce the number of orders entering the CQM formulation, and consequently, the quantum computer, thereby indirectly decreasing the complexity of the problem addressed by \texttt{LeapCQMHybrid}. It is also noteworthy that the creation of $P'$ also involves the creation of other subsets such as $M'$, $LT'$ or $UT'$. 

Ultimately, the development of \texttt{order\_filtering} results in a modification of step \textit{S1. Routing Problem Setup}, as shown in Fig. \ref{fig:Q4RPD}, with its updated version illustrated in Fig. \ref{fig:newS1}. By integrating this filtering mechanism, the system ensures that only reachable orders are included in the CQM formulation.

\section{Demonstration}\label{sec:demo}

In this section, we illustrate the practical application of \texttt{Q4RPD} highlighting the novel functionalities introduced in this paper. To solve the routing problem, as previously mentioned, it must be formulated as a CQM and then processed using D-Wave's \texttt{LeapCQMHybrid}. This solver is integrated into the \textit{Hybrid Solver Service} (HSS, \cite{HSS}), a collection of algorithms crafted by D-Wave. Specifically, the \texttt{LeapCQMHybrid} is divided into three distinct phases:

\begin{itemize}
    \item Initially, the CQM of the problem is input into a classical front-end, and a predetermined (but not configurable) number of equally structured threads is generated.
    
    \item Next, each created branch runs in parallel. Each thread consists of a Classical Heuristic Module (CM) and a Quantum Module (QM). The CM explores the solution space of the problem via classical heuristics, formulating various quantum queries to execute in the QM. These queries are partial representations of the problem tailored to the QPU capacity. The QPU's solutions are used to guide the CM towards promising areas of the solution search space. For the quantum queries, the \texttt{Advantage\_system7.1} is used, comprising 5616 qubits arranged in a Pegasus topology~\cite{boothby2020next}.
    
    \item Upon reaching the predefined time limit $T$ (set as 5 seconds in this study), all generated threads finish their execution and submit their solutions to the front-end. Then, \texttt{LeapCQMHybrid} selects and forwards the optimal solution identified among all threads. 
\end{itemize}

The use cases depicted along this section relate with \textit{simultaneous pickup and delivery}, \textit{time windows} and \textit{mobility restrictions by vehicle type}, respectively. To guarantee comprehensive evaluation, we assess \texttt{Q4RPD} with two extra instances, simultaneously integrating all the features outlined in this study. To facilitate replication, all instances and results are publicly accessible in \cite{PDPData}.

\begin{figure}[t!]
    \centering
    \includegraphics[width=0.77\linewidth]{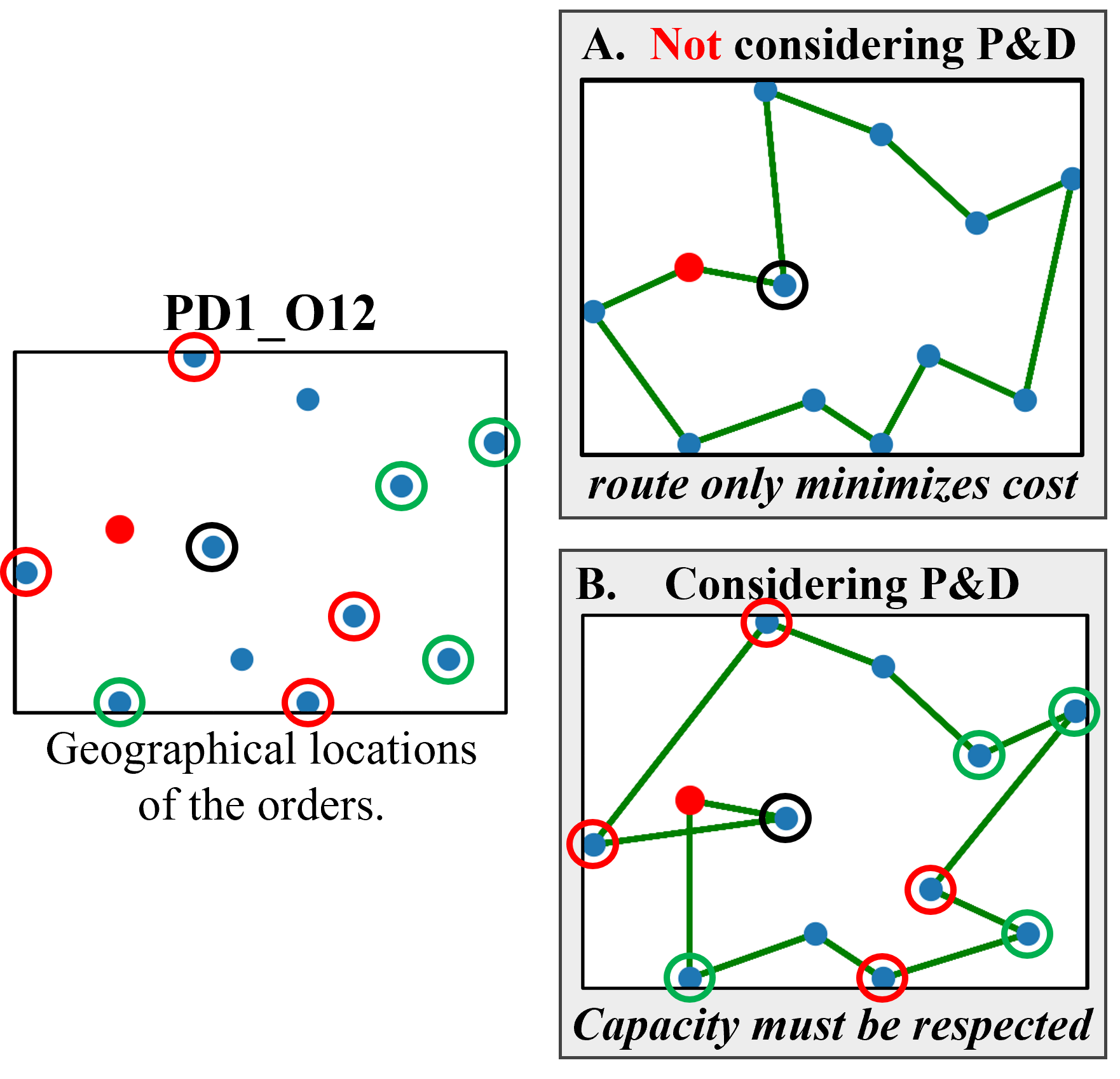}
    \caption{A 12-node PDP instance. A single truck with $W$=$D$=110 can complete the instance. For demonstration purposes, the black (\crule[Black]) rounded node should be visited last prior to the depot (which is the red node). The green (\crule[GreenLight]) rounded points indicate orders where the pickup demand is less than the delivery, creating space in the truck. The red (\crule[Red]) rounded nodes require more pickup than delivery, thus needing space in the truck. Two different executions are shown for demonstrating the impact of considering pickups and deliveries. In route \textbf{A}, pickups are not considered, so $wd_i$=$dd_i$=10 and $wp_i$=$dp_i$=0 in all $i$. In route \textbf{B}, for the red rounded nodes, $wd_i$=$dd_i$=5 and $wp_i$=$dp_i$=15, and for the green rounded nodes, $wd_i$=$dd_i$=15 and $wp_i$=$dp_i$=5. For the remaining orders, $wd_i$=$dd_i$=$wp_i$=$dp_i$=10. Neither time windows nor mobility restrictions are considered.}
    \label{fig:pd1}
\end{figure}

\subsection{Simultaneous Pickup and Delivery}\label{sec:hete}
\begin{figure}[t!]
    \centering
    \includegraphics[width=0.8\linewidth]{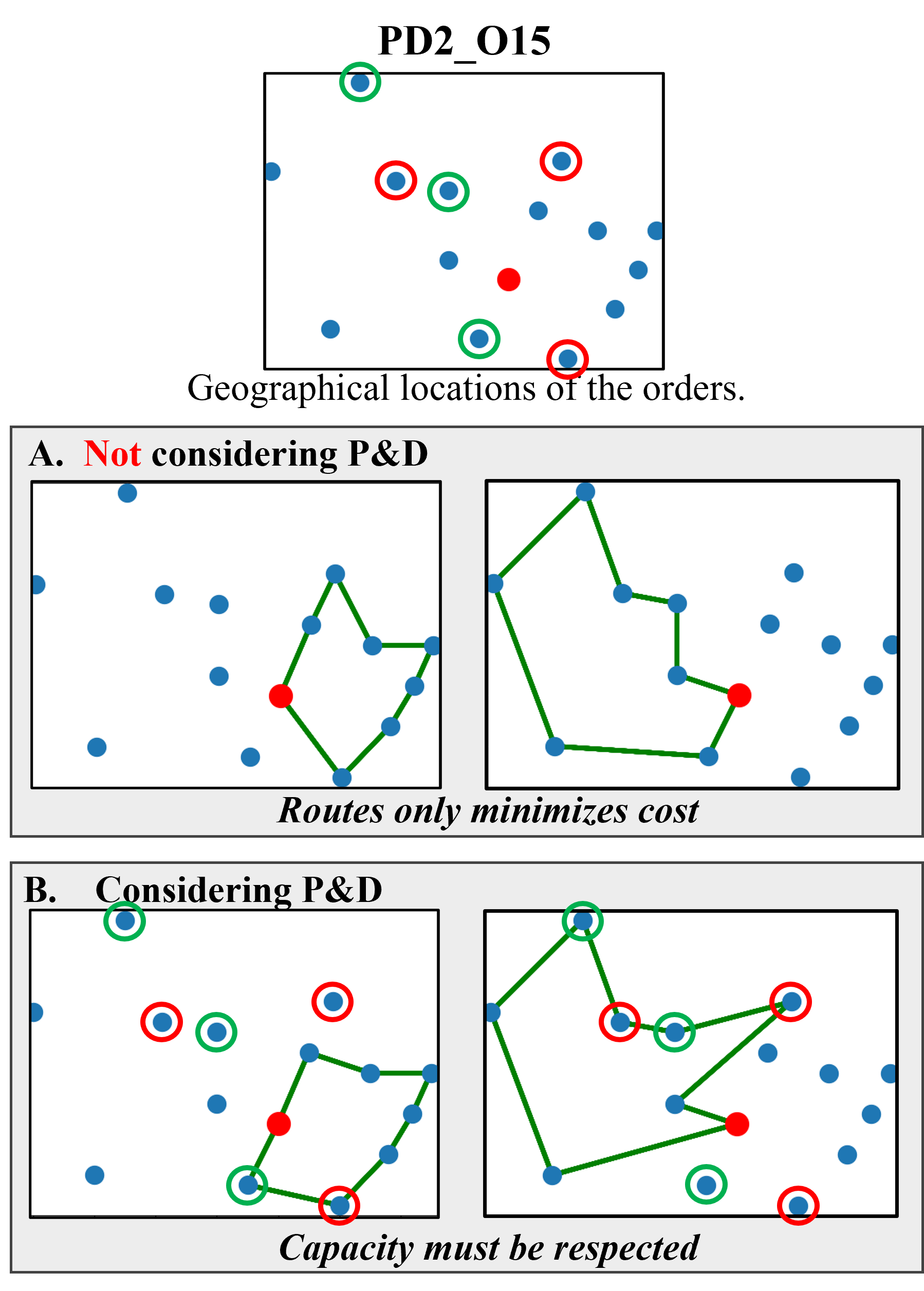}
    \caption{A 15-node PDP instance. Two identical trucks with $W = D = 70$ can complete the instance through two routes. The green (\crule[GreenLight]) and red (\crule[Red]) rounded points are configured in the same way as explained in the caption of Figure \ref{fig:pd1}. Solutions \textbf{A} and \textbf{B} are also calculated using the same procedure as in Fig. \ref{fig:pd1}. Neither time windows nor mobility restrictions are considered.}
    \label{fig:pd2}
\end{figure}

In this section, we illustrate how \texttt{Q4RPD} can manage simultaneous pickup and delivery restrictions. We use two instances for this purpose. The first example shows the impact of such a constraint on the formation of a single route, while the second example highlights the effect on a system composed of two distinct routes. Thus, Fig. \ref{fig:pd1} and Fig. \ref{fig:pd2} present the results of the examples related to this restriction, with further details provided in their captions.


It is crucial to emphasize that this restriction adds a layer of complexity to the problem. In logistic scenarios involving only deliveries, the adding of points to the route consistently approaches the solution to the boundaries set by capacity restrictions. Incorporating pickups into the problem means that each decision involves moving closer to or further away from the capacity limits. As a results, the solution space of the problem expands significantly. 

For instance, it might be preferable to first visit one or two distant points to deliver some packages, thereby increasing the truck's available capacity, and then pick up a package at a closer point. Such routes, which may resemble zigzags or returns in the opposite direction, would not be optimal in a routing problem focused solely on deliveries. This situation is illustrated in path B of Fig. \ref{fig:pd1}. This highlights the increased complexity of the new problem.

\subsection{Time Windows}\label{sec:TW}

\begin{figure}[b!]
    \centering
    \includegraphics[width=0.9\linewidth]{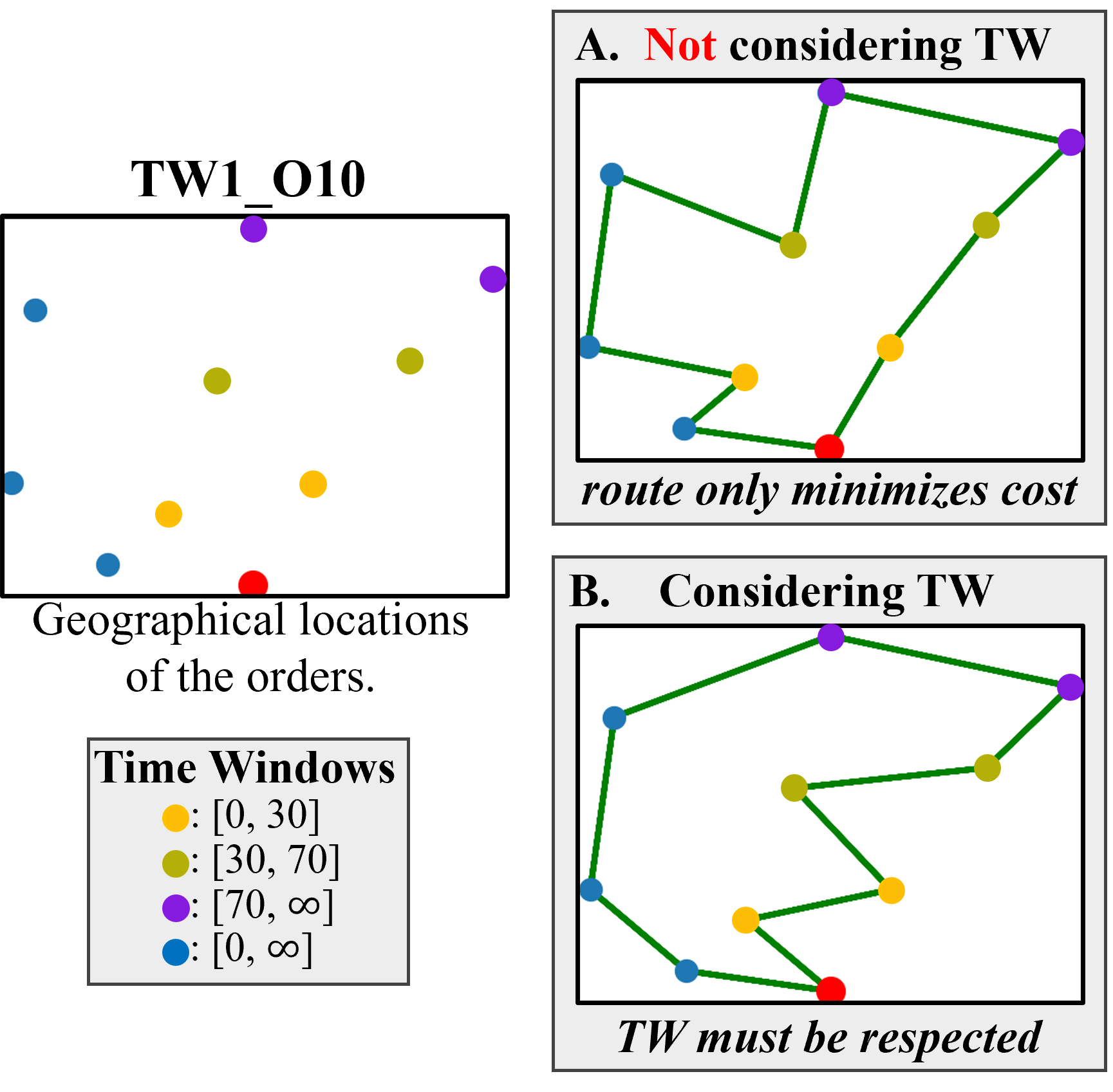}
    \caption{A 10-node PDP instance. In route \textbf{A}, time windows are not considered. In route \textbf{B}, the time windows [$lt_i,ut_i$] depicted in the lower-left frame are used. A single truck with $W$=$D$=110 can complete the instance at once. For all orders $wd_i$=$dd_i$=$wp_i$=$dp_i$=10. Neither mobility restrictions nor pickups are considered.}
    \label{fig:tw1}
\end{figure}

\begin{figure}[b!]
    \centering
    \includegraphics[width=0.85\linewidth]{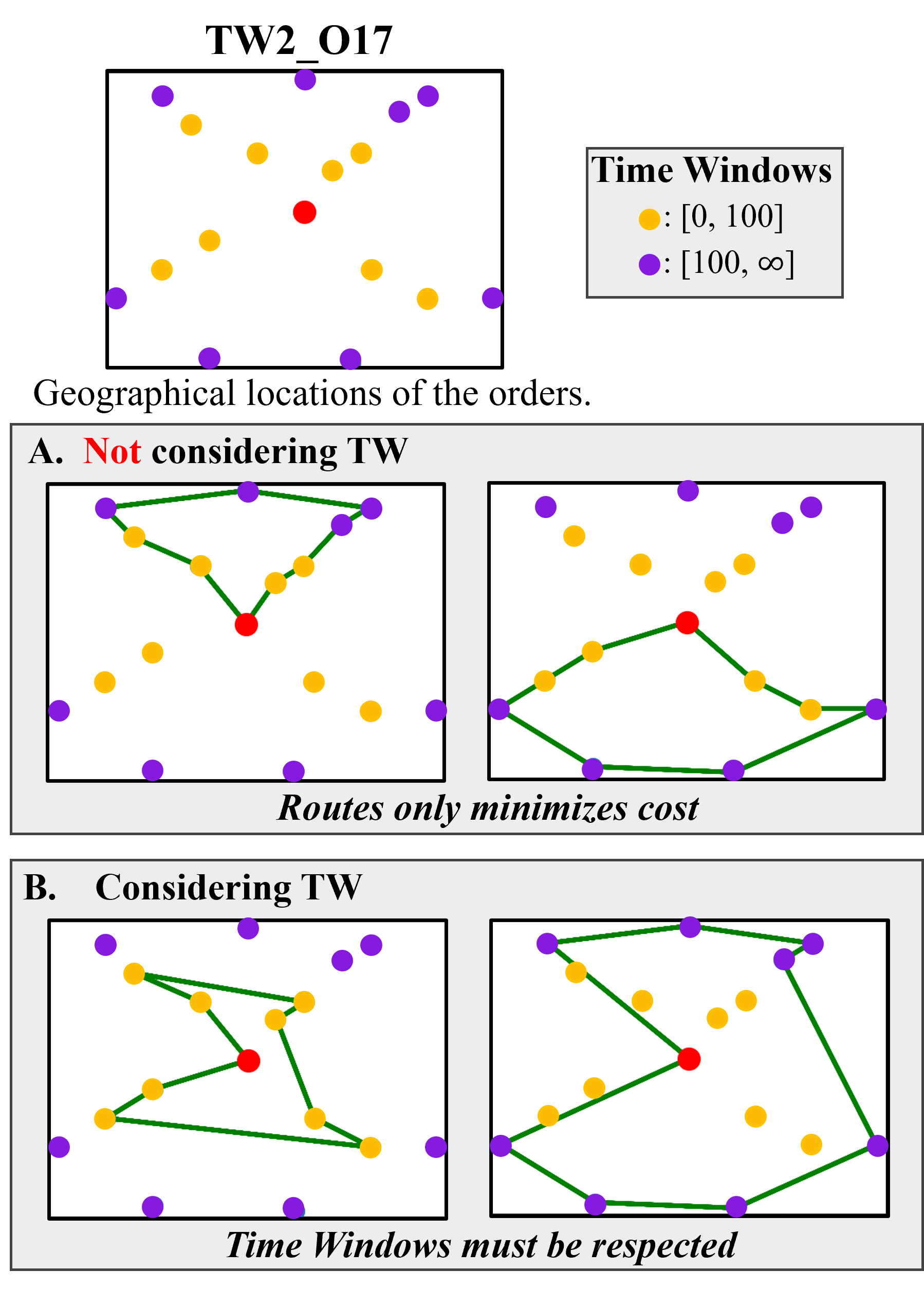}
    \caption{A 17-node PDP instance. In route \textbf{A}, time windows are not considered. In route \textbf{B}, the time windows [$lt_i,ut_i$] depicted in the upper-right frame are used. This scenario models a situation where orders are split into two categories: those that need to be executed in the morning and those that must be completed in the afternoon. Two identical trucks with $W = D = 70$ can complete the instance through two routes. For all orders $wd_i$=$dd_i$=$wp_i$=$dp_i$=10. Neither mobility restrictions nor pickups are considered.}
    \label{fig:tw2}
\end{figure}

In this subsection, we demonstrate how \texttt{Q4RPD} handles orders with designated time windows. We employ two distinct instances to examine the effect of this feature on a single route and in a scenario involving two paths. The outcomes for these instances are depicted in Fig. \ref{fig:tw1} and Fig. \ref{fig:tw2}, where four order types are distinguished based on the assigned window type. Different colors are used to differentiate the order types:

\begin{itemize}
    \item \textit{Purple nodes} (\crule[Purple]): for demands with just a lower time limit of $t'$, i.e., $lt_i = t',ut_i=\infty$ .
    \item \textit{Yellow nodes} (\crule[Yellow]): for orders with an upper time limit of $t''$, that is, $lt_i = 0,ut_i=t''$.
    \item \textit{Green nodes} (\crule[Green]): for depicting cases with both lower and upper time limits, i.e., $lt_i = t',ut_i=t''$.
    \item \textit{Blue nodes} (\crule[LightBlue]): for representing orders with no time window, that is, $lt_i = 0,ut_i = \infty$.
\end{itemize}

It is important to mention that, in this study, we consider both $t'$ and $t''$ as variables independent of real time. In other words, they represent the time the truck spends on the route. These values can later be assigned to real time by the company responsible for executing the routes.

\subsection{Mobility Restrictions by Vehicle Type}\label{sec:MR}
\begin{figure}[b!]
    \centering
    \includegraphics[width=0.9\linewidth]{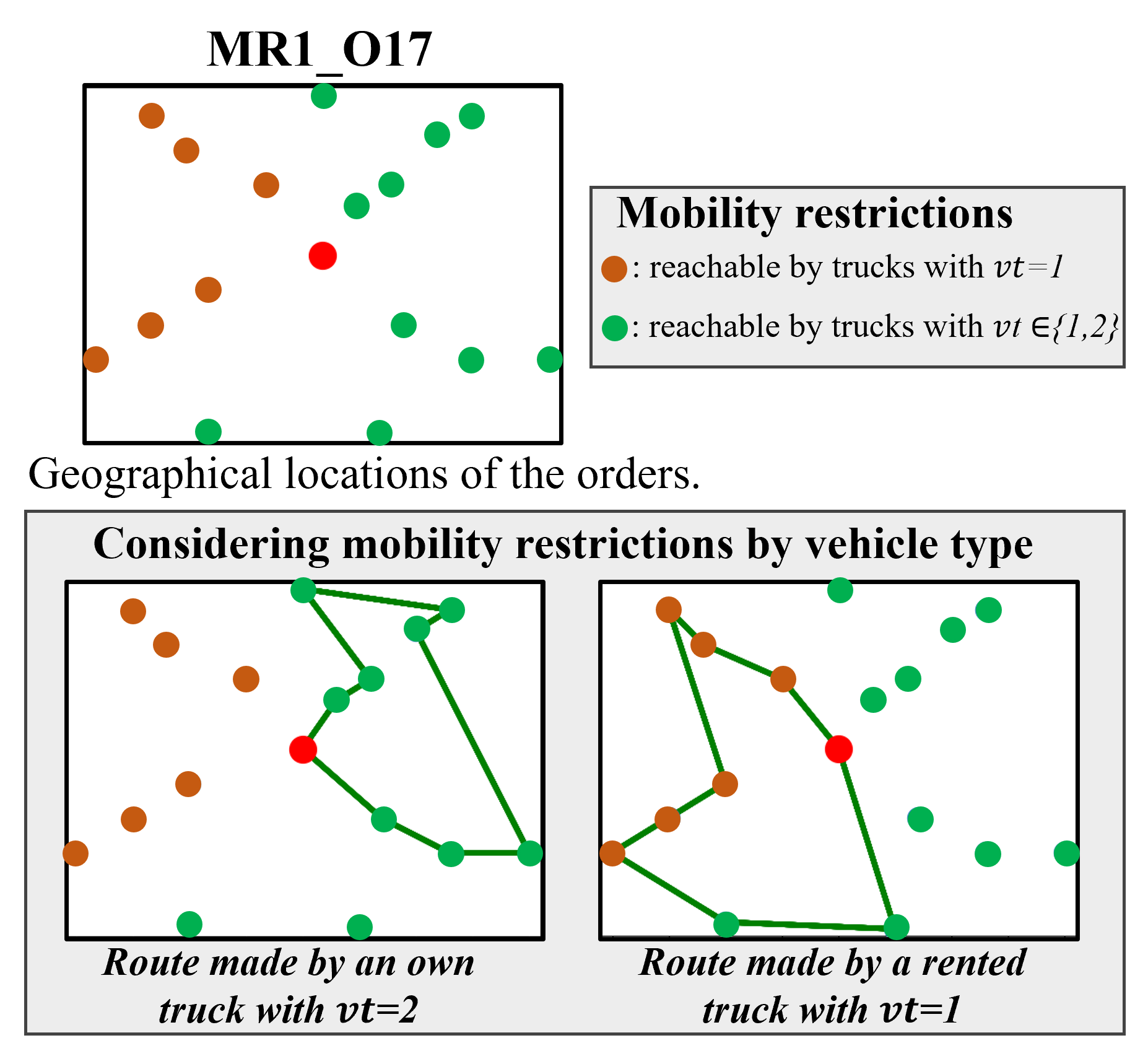}
    \caption{A 17-node PDP instance. This example models a situation where orders are split into two categories: those that can be served by trucks with $vt \in \{1,2\}$ (\crule[Green2]), and those that can only be reachable by vehicles with $vt=1$ (\crule[Brown]). For both trucks: $W = D = 70$. For all orders: $wd_i$=$dd_i$=$wp_i$=$dp_i$=10. Neither time windows nor pickups are considered.}
    \label{fig:mr}
\end{figure}

In this subsection, we employ a modified version of the \texttt{PD17\_P0} instance tailored to the mobility restrictions feature. Specifically, of the 16 orders to be completed, six are situated in an area accessible only to trucks with $vt=1$, while the remaining orders can be serviced by types with $vt \in \{1,2\}$. Given that the vehicle fleet includes only one owned truck, with $vt=2$, the user must rent an additional truck of $vt=1$ to satisfy all the orders. The results and specifics of this use case are detailed in the caption of Fig. \ref{fig:mr}.

\begin{figure*}[h!]
    \centering
    \includegraphics[width=0.95\linewidth]{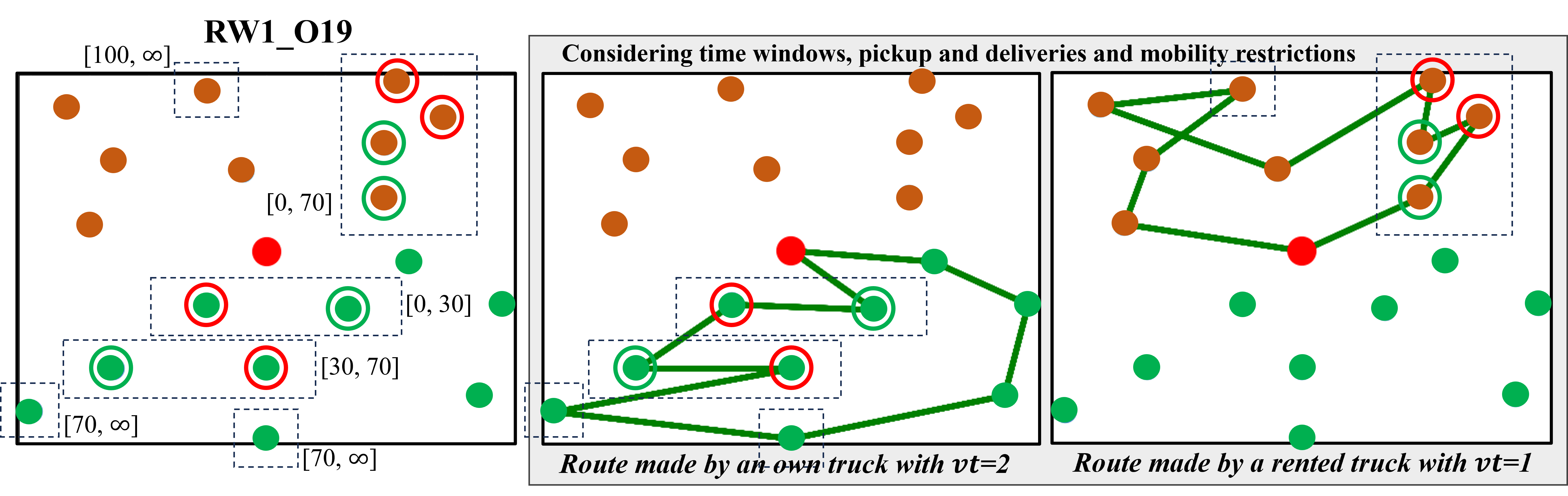}
    \caption{A 19-noded PDP instance. This example models a situation where orders are split into two categories: those that can be served by trucks with $vt \in \{1,2\}$ (\crule[Green2]), and those that can only be reachable by vehicles with $vt=1$ (\crule[Brown]). The green (\crule[GreenLight]) rounded points indicate orders where the pickup demand is less than the delivery, augmenting the capacity of the truck. The red (\crule[Red]) rounded nodes require more pickup than delivery, thus lacking space in the truck. Time windows are represented as [$lt_i,ut_i$], applying to all nodes within the squares of dashed points. We refer interested readers to \cite{PDPData}  for specific details on truck capacities and the parameters $wd_i$, $dd_i$, $wp_i$ and $dp_i$.}
    \label{fig:rw1}
\end{figure*}

\begin{figure*}[h!]
    \centering
    \includegraphics[width=1.0\linewidth]{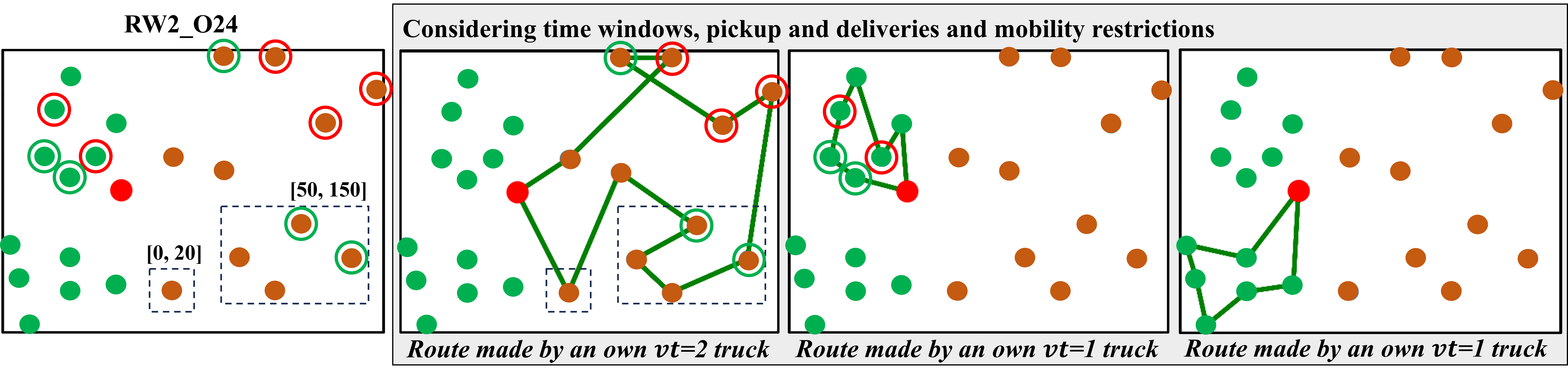}
    \caption{A 24-noded PDP instance. All visual characteristics of this instance have been configured similarly to those in Fig. \ref{fig:rw1}. We recommend readers refer to the caption of that figure for further details.}
    \label{fig:rw2}
\end{figure*}

\subsection{Advanced instances}\label{sec:adv}

In this final subsection, we evaluate the \texttt{Q4RPD} using two instances that incorporate all the constraints outlined in this research. The aim of this experimentation is to demonstrate that the \texttt{Q4RPD} can manage scenarios that closely resemble real-world environments (despite acknowledging that the real world is even more complex), with computational complexity and sizes competing with the current state of the art. With this in mind, the main characteristics of the two instances considered in this final test are as follows:

\begin{itemize}
    \item \texttt{RW1\_O19}: This 19-node instance is addressed by calculating two routes, each executed by trucks of different categories. This use case is designed to be the most restrictive in terms of time windows and pickups and deliveries, compelling the vehicles to follow counterintuitive routes to meet all imposed requirements.
    \item \texttt{RW2\_O24}: this 24-node instance is the largest addressed in this work. In this use case, we intend to simulate a scenario where a company operates its own fleet of vehicles, including a large-capacity truck with mobility restrictions and a smaller truck without such restrictions.
\end{itemize}

Additional details can be found in the captions of both images. For a more comprehensive understanding, we recommend readers access the public repository \cite{PDPData}, where all the instances discussed in this article, as well as all demonstrated results, are available.

\section{Conclusions and further work}\label{sec:conc}

This research has expanded upon our previous study \cite{osaba2024solving} by introducing new logistic applications. The advancements detailed in this paper have enabled our tool to handle more logistics-oriented routing scenarios. Concretely, we have detailed and tested three distinct features: \textit{i)} simultaneous pickups and deliveries, \textit{ii)} time-windows, and \textit{iii)} mobility restrictions by vehicle type. As future work, we intend to implement further functionalities such as forbidden paths or multiple depots. Furthermore, we plan to expand the application of \texttt{Q4RPD} to additional stages within the logistics distribution chain. This includes considering the hybridization of the system with methods designed to address other problems, such as the bin packing problem \cite{v2023hybrid}. We will also focus on enhancing the overall quality of the solutions given by \texttt{Q4RPD}. Additionally, we plan to explore the introduction of advanced paradigms such as transfer optimization in our future work.

\bibliographystyle{IEEEtran}
\bibliography{IEEEexample}

\begin{thebibliography}{10}
\providecommand{\url}[1]{#1}
\csname url@samestyle\endcsname
\providecommand{\newblock}{\relax}
\providecommand{\bibinfo}[2]{#2}
\providecommand{\BIBentrySTDinterwordspacing}{\spaceskip=0pt\relax}
\providecommand{\BIBentryALTinterwordstretchfactor}{4}
\providecommand{\BIBentryALTinterwordspacing}{\spaceskip=\fontdimen2\font plus
\BIBentryALTinterwordstretchfactor\fontdimen3\font minus
  \fontdimen4\font\relax}
\providecommand{\BIBforeignlanguage}[2]{{%
\expandafter\ifx\csname l@#1\endcsname\relax
\typeout{** WARNING: IEEEtran.bst: No hyphenation pattern has been}%
\typeout{** loaded for the language `#1'. Using the pattern for}%
\typeout{** the default language instead.}%
\else
\language=\csname l@#1\endcsname
\fi
#2}}
\providecommand{\BIBdecl}{\relax}
\BIBdecl

\bibitem{hanif2015applying}
S.~Hanif and T.~Holvoet, ``Applying delegate mas patterns in designing
  solutions for dynamic pickup and delivery problems,'' in \emph{Advances in
  Artificial Transportation Systems and Simulation}.\hskip 1em plus 0.5em minus
  0.4em\relax Elsevier, 2015, pp. 79--102.

\bibitem{benarbia2021literature}
T.~Benarbia and K.~Kyamakya, ``A literature review of drone-based package
  delivery logistics systems and their implementation feasibility,''
  \emph{Sustainability}, vol.~14, no.~1, p. 360, 2021.

\bibitem{osaba2022systematic}
E.~Osaba, E.~Villar-Rodriguez, and I.~Oregi, ``A systematic literature review
  of quantum computing for routing problems,'' \emph{IEEE Access}, vol.~10, pp.
  55\,805--55\,817, 2022.

\bibitem{gutin2006traveling}
G.~Gutin and A.~P. Punnen, \emph{The traveling salesman problem and its
  variations}.\hskip 1em plus 0.5em minus 0.4em\relax Springer Science \&
  Business Media, 2006, vol.~12.

\bibitem{toth2002vehicle}
P.~Toth and D.~Vigo, \emph{The vehicle routing problem}.\hskip 1em plus 0.5em
  minus 0.4em\relax SIAM, 2002.

\bibitem{osaba2024solving}
E.~Osaba, E.~Villar-Rodriguez, and A.~Asla, ``Solving a real-world package
  delivery routing problem using quantum annealers,'' \emph{Scientific
  Reports}, vol.~14, no.~1, p. 24791, 2024.

\bibitem{golden2008vehicle}
B.~L. Golden, S.~Raghavan, and E.~A. Wasil, \emph{The vehicle routing problem:
  latest advances and new challenges}.\hskip 1em plus 0.5em minus 0.4em\relax
  Springer Science \& Business Media, 2008, vol.~43.

\bibitem{dantzig1959truck}
G.~B. Dantzig and J.~H. Ramser, ``The truck dispatching problem,''
  \emph{Management science}, vol.~6, no.~1, pp. 80--91, 1959.

\bibitem{mor2022vehicle}
A.~Mor and M.~G. Speranza, ``Vehicle routing problems over time: a survey,''
  \emph{Annals of Operations Research}, vol. 314, no.~1, pp. 255--275, 2022.

\bibitem{EKSIOGLU20091472}
B.~Eksioglu, A.~V. Vural, and A.~Reisman, ``The vehicle routing problem: A
  taxonomic review,'' \emph{Computers \& Industrial Engineering}, vol.~57,
  no.~4, pp. 1472--1483, 2009.

\bibitem{braekers2016vehicle}
K.~Braekers, K.~Ramaekers, and I.~Van~Nieuwenhuyse, ``The vehicle routing
  problem: State of the art classification and review,'' \emph{Computers \&
  industrial engineering}, vol.~99, pp. 300--313, 2016.

\bibitem{chinan1}
\BIBentryALTinterwordspacing
Z.~Wang, R.~Zhu, J.-Y. Ding, Y.~Yang, and K.~You, ``An exact method for the
  daily package shipment problem with outsourcing,'' 2022. [Online]. Available:
  \url{https://arxiv.org/abs/2202.03614}
\BIBentrySTDinterwordspacing

\bibitem{heuristic2}
E.~Gussmagg-Pfliegl, F.~Tricoire, K.~F. Doerner, R.~F. Hartl, and S.~Irnich,
  ``Heuristics for a real-world mail delivery problem,'' in \emph{Applications
  of Evolutionary Computation: EvoApplications 2011: EvoCOMNET, EvoFIN, EvoHOT,
  EvoMUSART, EvoSTIM, and EvoTRANSLOG, Torino, Italy, April 27-29, 2011,
  Proceedings, Part II}.\hskip 1em plus 0.5em minus 0.4em\relax Springer, 2011,
  pp. 481--490.

\bibitem{parcel}
S.~Moganathan, S.~N. A.~M. Razali, N.~N.~H. Almaalei, and K.~Jacob,
  ``Comparison of metaheuristic approaches for parcel delivery problem,''
  \emph{International Journal of Logistics Systems and Management}, vol.~48,
  no.~1, pp. 67--91, 2024.

\bibitem{zhang2021solving}
Z.~Zhang, H.~Liu, M.~Zhou, and J.~Wang, ``Solving dynamic traveling salesman
  problems with deep reinforcement learning,'' \emph{IEEE Transactions on
  Neural Networks and Learning Systems}, vol.~34, no.~4, pp. 2119--2132, 2021.

\bibitem{xin2020step}
L.~Xin, W.~Song, Z.~Cao, and J.~Zhang, ``Step-wise deep learning models for
  solving routing problems,'' \emph{IEEE Transactions on Industrial
  Informatics}, vol.~17, no.~7, pp. 4861--4871, 2020.

\bibitem{altshuler2010anderson}
B.~Altshuler, H.~Krovi, and J.~Roland, ``Anderson localization makes adiabatic
  quantum optimization fail,'' \emph{Proceedings of the National Academy of
  Sciences}, vol. 107, no.~28, pp. 12\,446--12\,450, 2010.

\bibitem{somma2012quantum}
R.~D. Somma, D.~Nagaj, and M.~Kieferov{\'a}, ``Quantum speedup by quantum
  annealing,'' \emph{Physical review letters}, vol. 109, no.~5, p. 050501,
  2012.

\bibitem{hastings2021power}
M.~B. Hastings, ``The power of adiabatic quantum computation with no sign
  problem,'' \emph{Quantum}, vol.~5, p. 597, 2021.

\bibitem{tasseff2024emerging}
B.~Tasseff, T.~Albash, Z.~Morrell, M.~Vuffray, A.~Y. Lokhov, S.~Misra, and
  C.~Coffrin, ``On the emerging potential of quantum annealing hardware for
  combinatorial optimization,'' \emph{Journal of Heuristics}, vol.~30, no.~5,
  pp. 325--358, 2024.

\bibitem{azad2022solving}
U.~Azad, B.~K. Behera, E.~A. Ahmed, P.~K. Panigrahi, and A.~Farouk, ``Solving
  vehicle routing problem using quantum approximate optimization algorithm,''
  \emph{IEEE Transactions on Intelligent Transportation Systems}, vol.~24,
  no.~7, pp. 7564--7573, 2022.

\bibitem{mohanty2024solving}
N.~Mohanty, B.~K. Behera, and C.~Ferrie, ``Solving the vehicle routing problem
  via quantum support vector machines,'' \emph{Quantum Machine Intelligence},
  vol.~6, no.~1, p.~34, 2024.

\bibitem{tambunan2023quantum}
T.~D. Tambunan, A.~B. Suksmono, I.~J.~M. Edward, and R.~Mulyawan, ``Quantum
  annealing for vehicle routing problem with weighted segment,'' in \emph{AIP
  Conference Proceedings}, vol. 2906, no.~1.\hskip 1em plus 0.5em minus
  0.4em\relax AIP Publishing, 2023.

\bibitem{weinberg2023supply}
S.~J. Weinberg, F.~Sanches, T.~Ide, K.~Kamiya, and R.~Correll, ``Supply chain
  logistics with quantum and classical annealing algorithms,'' \emph{Scientific
  Reports}, vol.~13, no.~1, p. 4770, 2023.

\bibitem{cattelan2024modeling}
M.~Cattelan and S.~Yarkoni, ``Modeling routing problems in qubo with
  application to ride-hailing,'' \emph{Scientific Reports}, vol.~14, no.~1, p.
  19768, 2024.

\bibitem{leapCQM}
{D-Wave Developers}, ``{Measuring Performance of the Leap Constrained Quadratic
  Model Solver},'' D-Wave Systems Inc., Tech. Rep. 14-1065A-A, 11 2022.

\bibitem{phillipson2024quantum}
F.~Phillipson, ``Quantum computing in logistics and supply chain management-an
  overview,'' \emph{arXiv preprint arXiv:2402.17520}, 2024.

\bibitem{irie2019quantum}
H.~Irie, G.~Wongpaisarnsin, M.~Terabe, A.~Miki, and S.~Taguchi, ``Quantum
  annealing of vehicle routing problem with time, state and capacity,'' in
  \emph{Quantum Technology and Optimization Problems: First International
  Workshop, QTOP 2019, Munich, Germany, March 18, 2019, Proceedings 1}.\hskip
  1em plus 0.5em minus 0.4em\relax Springer, 2019, pp. 145--156.

\bibitem{harwood2021formulating}
S.~Harwood, C.~Gambella, D.~Trenev, A.~Simonetto, D.~Bernal, and D.~Greenberg,
  ``Formulating and solving routing problems on quantum computers,'' \emph{IEEE
  Transactions on Quantum Engineering}, vol.~2, pp. 1--17, 2021.

\bibitem{HSS}
{D-Wave Developers}, ``{D-Wave Hybrid Solver Service: An Overview},'' D-Wave
  Systems Inc., Tech. Rep. 14-1039A-B, 05 2020.

\bibitem{boothby2020next}
K.~Boothby, P.~Bunyk, J.~Raymond, and A.~Roy, ``Next-generation topology of
  d-wave quantum processors,'' \emph{arXiv preprint arXiv:2003.00133}, 2020.

\bibitem{PDPData}
E.~Osaba, P.~Miranda-Rodriguez, and E.~Villar, ``Benchmark dataset and results
  for a real-world package delivery problem,''
  \url{https://doi.org/10.17632/253wjnx2hp.1}, 2025, online at Mendeley Data.

\bibitem{v2023hybrid}
S.~V.~Romero, E.~Osaba, E.~Villar-Rodriguez, I.~Oregi, and Y.~Ban, ``Hybrid
  approach for solving real-world bin packing problem instances using quantum
  annealers,'' \emph{Scientific Reports}, vol.~13, no.~1, p. 11777, 2023.

\end{thebibliography}
\end{document}